\documentclass[aps,pra,superscriptaddress, twocolumn, preprintnumbers,amsmath,amssymb,nobalancelastpage,10pt,longbibliography]{revtex4-1}

\usepackage{amsmath}
\usepackage{verbatim}
\usepackage{graphicx}
\usepackage{color}
\usepackage{braket}
\usepackage{ bbold }
\usepackage{ bbold }
\usepackage{yfonts}
\usepackage[normalem]{ulem}

\usepackage{graphicx}
\usepackage{dcolumn}
\usepackage{bm}
\usepackage[dvipsnames]{xcolor}

\usepackage[colorlinks=true,citecolor=blue,linkcolor=blue,urlcolor=blue]{hyperref}
\usepackage{subfigure}
\definecolor{darkgreen}{rgb}{0,0.60,.2}

\begin{document}

\title{Localization and multifractal properties of the long-range Kitaev chain in the presence of an Aubry-Andr\'e-Harper modulation}

\author{Joana Fraxanet}
 \affiliation{ICFO - Institut de Ci\`encies Fot\`oniques, The Barcelona Institute of Science and Technology, 08860 Castelldefels (Barcelona), Spain}
\author{Utso Bhattacharya}
 \affiliation{ICFO - Institut de Ci\`encies Fot\`oniques, The Barcelona Institute of Science and Technology, 08860 Castelldefels (Barcelona), Spain}
\author{Tobias Grass}
 \affiliation{ICFO - Institut de Ci\`encies Fot\`oniques, The Barcelona Institute of Science and Technology, 08860 Castelldefels (Barcelona), Spain}
\author{Maciej Lewenstein}
 \affiliation{ICFO - Institut de Ci\`encies Fot\`oniques, The Barcelona Institute of Science and Technology, 08860 Castelldefels (Barcelona), Spain}
\affiliation{ICREA, Pg. Lluis Companys 23, 08010 Barcelona, Spain}
\author{Alexandre Dauphin}
 \affiliation{ICFO - Institut de Ci\`encies Fot\`oniques, The Barcelona Institute of Science and Technology, 08860 Castelldefels (Barcelona), Spain}

\date{\today}

\begin{abstract}
In the presence of quasi-periodic potentials, the celebrated Kitaev chain presents an intriguing phase diagram with ergodic, localized and and multifractal states. In this work, we generalize these results by studying the localization properties of the Aubry-Andr\'e-Harper model in the presence of long-range hopping and superconducting pairing amplitudes. These amplitudes decay with power-law exponents $\xi$ and $\alpha$ respectively. To this end, we review and compare a toolbox of global and local characterization methods in order to investigate different types of transitions between ergodic, localized and multifractal states. We report energy-dependent transitions from ergodic to multifractal states for pairing terms with $\alpha<1$ and energy-dependent transitions from ergodic to localized states with an intermediate multifractal region for $\alpha>1$. The size of the intermediate multifractal region depends not only on the value of the superconducting pairing term $\Delta$, but also on the energy band. The transitions are not described by a mobility edge, but instead we report hybridization of bands with different types of localization properties. This leads to coexisting multifractal regimes where fractal dimensions follow different distributions. 
\end{abstract}

\maketitle

\section{\label{sec:intro} Introduction}

Understanding metal-insulator transitions is one of the central questions in condensed matter physics. More specifically, there is a need for a better understanding and a precise characterization of the transition points, which are known to present critical properties. There are many examples of metal-insulator transitions in nature, but one of the most paradigmatic examples is Anderson localization~\cite{anderson58}, in which a system becomes insulating in the presence of disorder.

Disordered non-interacting models exhibiting Anderson localization represent a very interesting playground to investigate this type of transition, but the random nature of disorder requires averaging over many disorder realizations. Moreover, no metal-insulator transitions can be observed in one-dimensional models, where a system is an insulator for any strength of the disorder. In contrast, quasi-periodic systems have recently gained a lot of attention as an alternative to explore localization and criticality. In particular, such models are neither periodic nor disordered, but still they show non-trivial localization properties even for very simplistic Hamiltonians, and even in one spatial dimension.

One of the most well known examples featuring a metal-insulator transition in one dimension is the Aubry-Andr\'e-Harper (AAH) model~\cite{aubry80,harper55}, resulting from the superposition of two incommensurate lattices. Specifically, we refer to a model with nearest neighbor hopping and where one of the lattices is assumed to be weak and can be treated as a perturbation, leading to an incommensurate quasi-periodic potential. For certain values of such quasi-periodic potential, a transition between ergodic and localized states takes place~\cite{tang86,hiramoto89}. This transition can be determined from the self-duality property of the Hamiltonian at the critical point: the eigenfunctions have the same distribution in real and momentum space as the Schr\"odinger equation is equal to its Fourier transform. Exactly at the critical point all the states of the system are neither localized nor ergodic but multifractal: they are locally scale-invariant and their structure can be characterized using one or several non-integer fractal dimensions. The AAH model has been realized in several experimental setups including ultracold atoms in optical lattices~\cite{luschen18, modugno10,roati08,lohse16,nakajima16,fangzhao21} and photonic devices~\cite{tanese14}, and these realizations have helped to better characterize the localization transition. Moreover, the AAH model has also been extensively studied for its topological properties~\cite{kraus2012,kraus2012a,tanese2014,dareau2017,madsen13}. In fact, the AAH Hamiltonians can be seen as the dimensional reduction of a 2D Hofstadter model~\cite{hofstadter76}, which describes electrons in a 2D lattice subjected to a perpendicular magnetic field akin to that of a quantum Hall system. 

In order to discover and characterize new types of metal-insulator transitions, great effort has been devoted to studying different generalizations of this one-dimensional quasi-periodic system. Specifically, one can ask what happens in the presence of long-range processes and/or interactions \cite{maity2019,saha2019, saha2019b, saha2021,  mondal2021}. For example, a recent work~\cite{biddle2011} has studied the interplay between the quasi-periodicity and long-range hopping amplitudes. For exponentially decaying hoppings, such a system exhibits an energy dependent mobility edge which, in certain cases, can be predicted by generalizing the self-duality conditions. Specifically, this means that for a fixed Hamiltonian the system can have localized states for low energies and ergodic states for higher energies, with a mobility edge separating the two. On the other hand, power law decaying hopping amplitudes~\cite{deng2019} lead to a mobility edge that shows a block-like structure in terms of the quasi-periodic potential, which can be characterized through the mathematical Diophantine nature of the incommensurability~\cite{nilanjan21}, and slow decay of the hoppings leads to a new type of transition between ergodic and multifractal states. In addition, one can also study the effect of interactions. In particular, density-density interactions~\cite{cookmeyer20} or Hamiltonians with superconducting pairing terms~\cite{cai13, wang2016, zeng16, liu2017, wang2016b, yahyavi19,  liu2020, lv2022} have been considered, and lead to an intermediate extended multifractal regime, which exists for a finite range of quasi-periodic potential strengths. Other works have considered coupling of more than one Aubry-Andr\'e-Harper chains~\cite{rossignolo19, li2020}, or two-dimensional Aubry-Andr\'e-Harper models~\cite{scheider2020}, which leads to hybridization of ergodic and localized bands. There are also studies of more complicated quasi-periodic potentials \cite{liu2021, goblot2020, li2020, duthie21}, or systems of shallow lattices with incommensurability~\cite{yao19, biddle09}. In some cases, it is possible to analytically describe the transitions by generalizing the condition of self-duality of the system~\cite{sarang2017, biddle2011} or extracting information from the parent Hamiltonian~\cite{borgnia2021l, borgnia2021r}, but understanding the nature of the localization and multifractal properties is often challenging, and therefore, most of the recent works generally focus on numerical studies to characterize the transition points.

Both long-range Hamiltonians and Hamiltonians with superconducting pairing terms lead to interesting extended multifractal regions, which are difficult to characterize and whose critical properties are still not well understood. The combination of long-range pairing and quasi-periodicity has not yet been explored. In this work, we present a comprehensive study of the long-range Aubry-Andr\'e-Harper (AAH) model with superconducting pairing, generalizing the model presented in~\cite{fraxanet2020}. Using different methods for the characterization of localized, ergodic and multifractal regimes of the energy spectrum we show that long-range superconducting pairing creates energy-dependent transitions from ergodic to multifractal states for a slow decay of the power-law pairing. When the decay exponent is large, energy-dependent transitions occur from ergodic to localized states with an intermediate extended multifractal region. The intermediate multifractal region depends not only on the value of the superconducting pairing term $\Delta$, but also on the energy, and therefore the transitions cannot be described through a mobility edge. We instead report the hybridization of sets of energy bands with different type of localization. In particular,  this leads to coexisting multifractal regimes where fractal dimensions are distributed differently. 

\textbf{Plan of the paper.} In Section~\ref{sec:model}, we introduce the Hamiltonian and the specific limits that we will study. Then, in Section~\ref{sec:methods}, we present a toolbox of methods for the characterization of localized, ergodic and multifractal states, which we benchmark in two already known limits of the Hamiltonian. Finally, in section~\ref{sec:results}, we use this toolbox to characterize the effects of the combination of long-range hopping and superconducting pairing.

\section{\label{sec:model} The model}

Let us consider a tight-binding Hamiltonian of spinless fermions with long-range hopping and pairing terms, both following a power law, and an onsite quasi-periodic potential. The model is described by the Hamiltonian
\begin{equation}\label{eq: hamiltonian}
    H =-\sum_{i=0}^{N-1}\left(\sum_{l=1}^{N-1}\frac{t}{l^\xi} c^\dagger_{i+l}c_{i}  + \frac{\Delta}{l^\alpha} c^\dagger_{i+l}c^\dagger_{i}+\text{h.c.}\right)
    +2V f(i) c^\dagger_{i}c_{i}  
\end{equation}
where $t$ and $\Delta$ are the hopping and pairing amplitudes, $V$ is the quasi-periodic potential, and $c_{i}(c^\dagger_{i})$ is the annihilation (creation) operator at the $i$-th site of the chain. Both the hopping and the superconducting pairing terms follow a power law decay, depending on the distance $l$~\footnote{we here set the lattice spacing to unity} between the sites and the respective decay exponents $\alpha,\xi >0$. 

In the case of a constant quasi-periodic potential $f(i)=1$, the model has well known limits. 
For $\alpha,\xi \rightarrow \infty$, the system can be exactly mapped to the short-range Kitaev model with nearest-neighbor pairing terms~\cite{kitaev01}. In fact, even when the decay exponents are finite, as long as $\alpha, \xi > 1$ the system is still topologically equivalent to the short-range Kitaev model, and thus we will denote it as short-range power-law regime. In contrast, for $\alpha \ll 1$ and $\xi \gg 1$ the model is known to host a long-range topological phase with massive Dirac modes (MDM) characterized by a half integer topological invariant~\cite{bhattacharya19, vodola14, viyuela16, lepori17}, and for $\alpha \gg 1$ and $\xi \ll 1$, we recover a one-dimensional chain of spinless fermions with long-range hopping and short-range power-law pairing. Such regimes will be denoted as long-range superconducting pairing and long-range hoppings respectively.

In this work, we study the effect of the combination of the power-law superconducting pairing and an AAH modulation
\begin{equation}\label{eq:modulation}
f(i) = \cos (2 \pi \beta i + \phi )
\end{equation}
on the localization properties of the eigenstates. The modulation frequency $\beta$ gives the periodicity of the potential and the offset parameter $\phi$ shifts the origin of the modulation~\footnote{In this work, we fix the value of the offset parameter at $\phi = 4$. A non-zero value of the offset parameter is needed in order to avoid reflection symmetries in the system, which lead to a degenerate spectrum. Otherwise, the offset parameter value does not affect the multifractal and localization properties of the system.}. In particular, we take $\beta$ to be the inverse of the golden ratio $\tau^{-1} = \frac{\sqrt{5}-1}{2}$~\cite{aubry80}, which for finite systems of size $F_m$ can be approximated using the Fibonacci series as $\tau \approx F_{m-1}/F_m$~\footnote{The ratio of two consecutive Fibonacci numbers $\frac{F_{m-1}}{F_m}$, defined recursively as $F_0 = F_1 = 1$ and $F_{m+1} = F_{m}+F_{m-1}$, converges to the inverse of the golden ratio for $m \rightarrow \infty$. By choosing a length $N = F_m$ for a modulation frequency $\beta = \frac{F_{m-1}}{F_m}$, we make sure that the onsite quasi-periodic potential $f(i)$ is incommensurate within our system with respect to the lattice periodicity.}. For a nonzero AAH quasi-periodic potential and both $\xi, \alpha \rightarrow \infty$, we recover the short-range Kitaev model with an AAH modulation studied in Refs.~\cite{yahyavi19, wang2016, zeng16, lv2022}. For $\Delta = 0$ and $\xi \ll 1$, we retrieve the AAH model with long-range hopping amplitudes characterized in Refs.~\cite{deng2019, nilanjan21}.

\section{\label{sec:methods} Toolbox for the characterization of the localization of the eigenstates}

We now discuss a toolbox of methods used to characterize localized, ergodic and multifractal regions of the energy spectrum. In particular, we consider two classes of methods that we will denominate as global and local measures of localization and multifractality. The global methods characterize properties of all the eigenstates of the Hamiltonian. They are particularly useful when there is a simultaneous transition of all the eigenstates between, for example, ergodic and localized phases. The local methods characterize localization properties of few eigenstates or a band of eigenstates and are particularly useful in the presence of a mobility edge.

\subsection{Global characterization}\label{sub:gmethods}

\begin{figure}[t]
\includegraphics[width=\columnwidth]{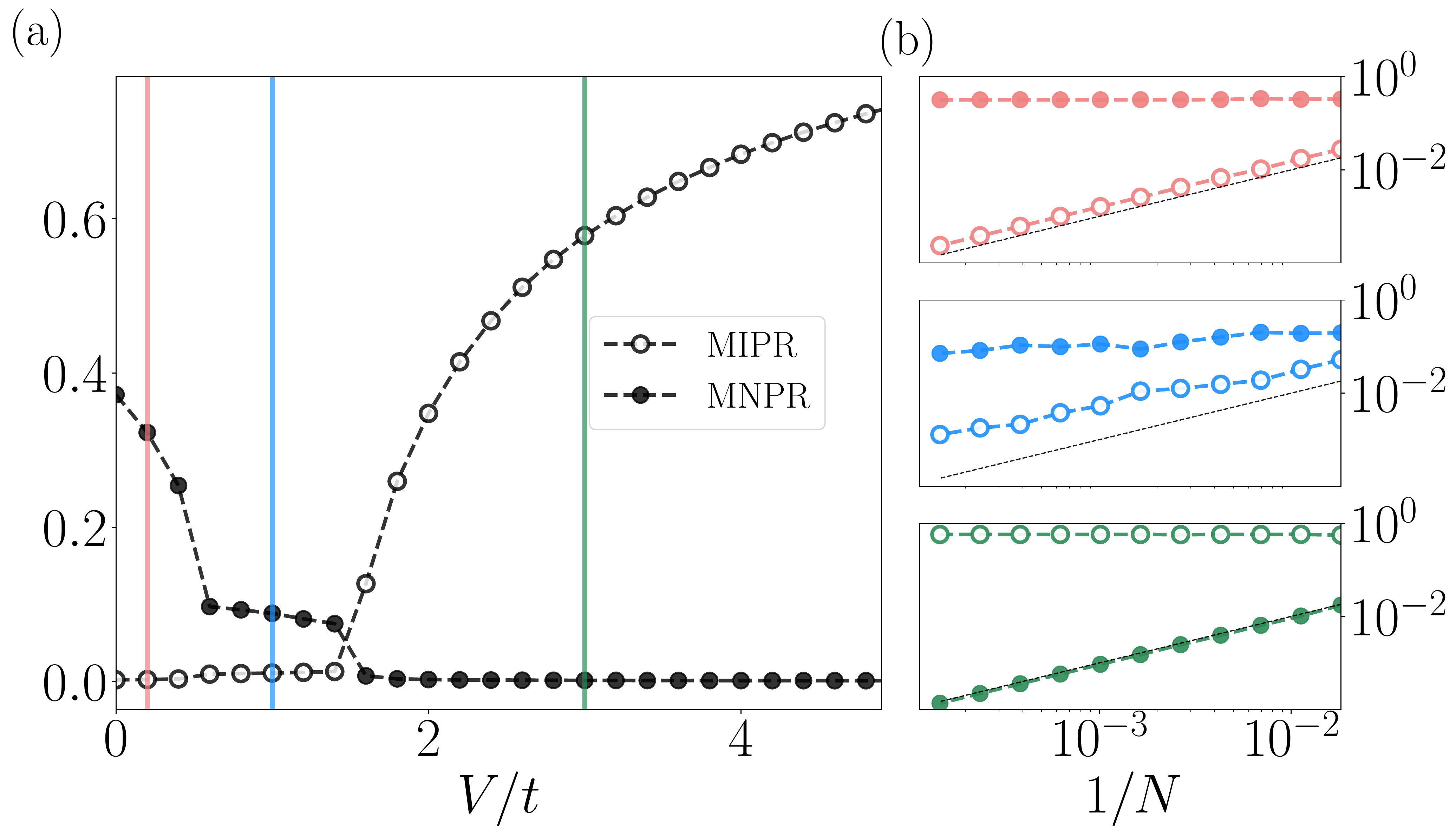}
\caption{\label{fig:gmeasures} \textbf{Mean inverse and normalized participation ratios.} \textbf{(a)} MIPR and MNPR for the Hamiltonian in eq.~\eqref{eq: hamiltonian_SR_AAH} with $N = 1597$ and $\Delta/t = 0.5$. We see three different regions corresponding to an ergodic, multifractal and localized spectrum respectively. The vertical lines correspond to $V/t = 0.2$ (red), $V/t = 1$ (blue) and $V/t = 3$ (green). \textbf{(b)} Finite size scaling of the MIPR (open circles) and MNPR (filled circles) at the indicated points in (a) for different values of $N$. For ergodic states (red) the MNPR converges to a nonzero value while the MIPR decays as following $N^{-1}$ (black dashed line). For localized states (green) it is the opposite. For multifractal states (blue), neither the MIPR nor the MNPR decay as $N^{-1}$.}
\end{figure}

\subsubsection{Short-range Kitaev limit} 
We illustrate the use of the global methods by studying the Hamiltonian~\eqref{eq: hamiltonian} in the short-range limit $\xi, \alpha \rightarrow \infty$
\begin{equation}\label{eq: hamiltonian_SR_AAH}
    H_{\xi, \alpha \rightarrow \infty} =-\sum_{i=0}^{N-1}\left(t c^\dagger_{i+1}c_{i}  + \Delta c^\dagger_{i+1}c^\dagger_{i}+\text{h.c.}\right)
    +2V f(i) c^\dagger_{i}c_{i}.  
\end{equation}
We here fix $\Delta/t = 0.5$ and then solve the eigenvalue problem using exact diagonalization for different values of the on-site quasi-periodic potential $V$.  We write the Hamiltonian in the Bogoliubov-de Gennes (BdG)~\cite{kitaev01} basis to properly treat the superconducting pairing term and consider periodic boundary conditions (PBC). For small values of $V\ll t$, the totality of the states of the system are ergodic, while for large values of $V \gg t$, all the eigenvectors are localized. In between, in contrast to the conventional Aubry-Andr\'e-Harper model, which includes no pairing terms and features a direct transition between ergodic and localized states, this system also has a multifractal region for $V \in \left[\vert \Delta - t\vert,\vert \Delta + t\vert\right]$~\cite{yahyavi19, wang2016, zeng16, lv2022}. In this Section, we characterize the phase diagram of the Hamiltonian $H_{\xi, \alpha \rightarrow \infty}$ with the help of two global observables: the mean participation ratios and the mean fractal dimension $D_2$. 


\subsubsection{Mean participation ratios} \label{subsub:MIPR}

The mean inverse participation ratio (MIPR) is one of the most used quantities to characterize the localization of the eigenstates. For systems with superconducting pairing terms, it can be defined as follows
\begin{equation}\label{eq:MIPR}
    \text{MIPR} = \frac{1}{2N}\sum_{n=1}^{2N} \sum_{j=1}^N (\vert u_{n,j}\vert^2 + \vert v_{n,j}\vert^2)^2,
\end{equation}

where  $n$ is the index of the eigenstate and $u_{n,j}$ and $v_{n,j}$ are the coefficients of the $n$th eigenvector for a site $j$ in the Bogoliubov de Gennes (BdG) basis, corresponding to $c_j$ and $c^\dagger_j$ respectively~\footnote{Note that we work in the BdG basis and, therefore, we have $2N$ eigenstates}. Note that, in contrast to~\cite{yahyavi19, wang2016}, we take the square of the onsite occupation $ p_{n,j} =\vert u_{n,j}\vert^2+\vert v_{n,j}\vert^2$ defined on site $j$ for an eigenstate $n$. From the definition of the MIPR in eq.~\eqref{eq:MIPR}, we can generalize the mean normalized participation ratio (MNPR) defined in~\cite{li2020} for states in the BdG basis. Then, the MNPR reads 
\begin{equation}
    \text{MNPR} = \frac{1}{2N}\sum_{n=1}^{2N} \left(2N \sum_{j=1}^N (\vert u_{n,j}\vert^2 + \vert v_{n,j}\vert^2)^2\right)^{-1}.
\end{equation}
This quantity also represents a very useful quantity to probe the localization of the eigenstates. In particular, we emphasize that, for normalized eigenstates, which satisfy
\begin{equation}
    \sum_{n=1}^{N} (\vert u_{n,j}\vert^2 + \vert v_{n,j}\vert^2) = 1,
\end{equation}
the MIPR and the MNPR always lie between zero and one.

Figure~\ref{fig:gmeasures}(a) shows both the MIPR and the MNPR for the Hamiltonian $H_{\xi, \alpha \rightarrow \infty}$ and for different values of $V$. We now show how the combination of these two quantities already allows us to distinguish between localized states, ergodic states and multifractal states. The transitions can qualitatively be located at $V/t \approx 0.5$ and $V/t \approx 1.5$, even though the MIPR does not allow for a clear distinction between ergodic and multifractal states. In order to properly distinguish between the three types of states, we study the scaling of these quantities. Figure~\ref{fig:gmeasures}(b) shows the scaling properties of the MIPR and MNPR for $V/t= \{0.2,1,3\}$. For $V/t = 0.2$ (red), the eigenstates are ergodic, and the MNPR is finite within the thermodynamic limit, while the MIPR decays with the size of the system as $N^{-1}$ (black dashed line). This is because for an ergodic state, i.e. a completely delocalized state, the probability density is distributed uniformly and at each site it will be proportional to $N^{-1}$. For $V/t = 1$ (blue), the states are multifractal. Both the MNPR and MIPR depend on the system size and, most importantly, their decay deviates from the $N^{-1}$ behavior. For $V/t = 3$ (green), the states are localized. Then, the MIPR converges to a finite value while the MNPR decays as $N^{-1}$, as a consequence of having a probability density which vanishes everywhere but in one site for completely localized states. 

\subsubsection{Multifractal analysis of the wavefunctions}

We now focus on the multifractal regime with finite MIPR and MNPR. A multifractal system is a generalization of a fractal system in which several fractal dimensions are needed to fully describe its structure~\cite{salat2016}. Therefore, the full picture of a multifractal state cannot be reflected in quantities such as the mean participation ratio, and instead one needs to compute a continuous spectrum of scalings~\cite{hiramoto89}. Even though in this work we are mainly interested in distinguishing between ergodic, multifractal and localized regions of the phase diagram, we believe that it is also important to briefly introduce how to rigorously characterize the multifractality of the eigenstates~\cite{liu2017, kohmoto08} and also explain what is the difference between fractal and multifractal structures.

\begin{figure}[ht]
\includegraphics[width=\columnwidth]{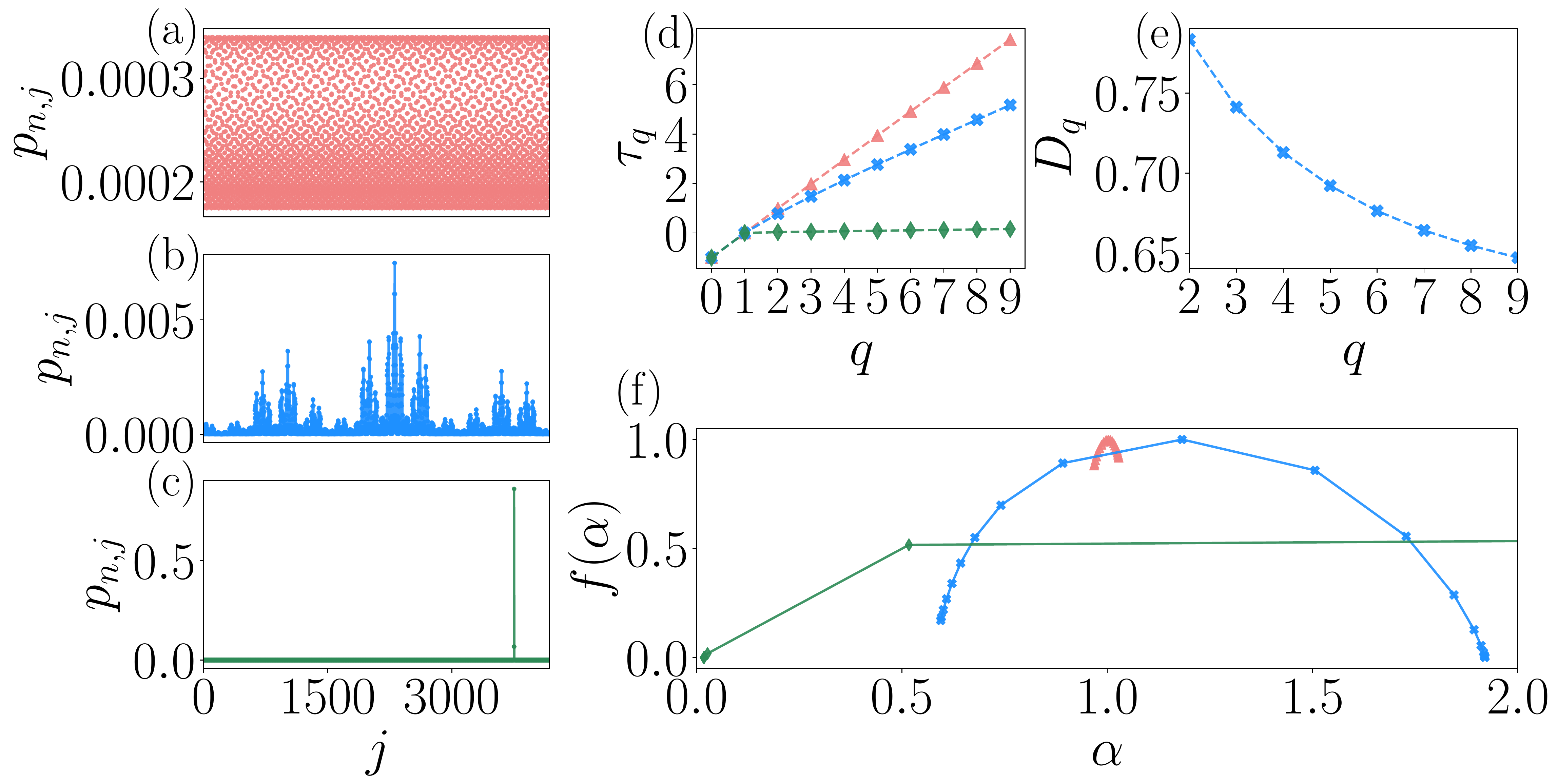}
\caption{\label{fig:mf} \textbf{Mean multifractal scaling of the wavefunctions.} \textbf{(a-c)} Onsite occupation $p_{n,j}$ of the ground state ($n=0$) of an ergodic system with $V/t = 0.2$ (a), multifractal system with $V/t = 1$ (b) and localized system with $V/t = 3$ (c). \textbf{(d)} $\tau_q$ as a function of the moments $q$ for an ergodic (red), localized (green) and multifractal (blue) ground states. \textbf{(e)} Generalized $q$-fractal dimensions for a multifractal groundstate. \textbf{(f)} Multifractal spectrum of an ergodic (red), localized (green) and multifractal (blue). For this calculation, we have taken integer values of $q\in [-10,10]$. For (a-f) we take the Hamiltonian in eq.~\eqref{eq: hamiltonian_SR_AAH} and consider $N = 4181$ and $\Delta/t = 0.5$. }
\end{figure}

Let us consider again the model in Eq.~\eqref{eq: hamiltonian_SR_AAH} with fixed size $N$ and select a given normalized wavefunction, which in this case will be the ground state ($n=0$). Recall that we define the onsite occupation as $ p_{n,j} =\vert u_{n,j}\vert^2+\vert v_{n,j}\vert^2$. We can generalize the notion of IPR by considering the following relation~\cite{roy2018}
\begin{equation}
 	\sum_{j=1}^N (p_{0,j})^q \sim N^{-\tau_q},
\end{equation}
where $\tau_q = D_q (q-1)$. Here the moment $q$ is an integer and $D_q$ for $q>1$ is known as the $q$-fractal dimension~\cite{mace17, evers08, deng2019}. For ergodic or localized states, we expect $D_q$ to be $1$ or $0$ respectively, regardless of the value of $q>1$ that we consider. This is because for an ergodic state, depicted in Fig.~\ref{fig:mf} for $V/t = 0.2$ (a), the probability density is distributed uniformly, all sites show a similar scaling that is propotional to $1/N$ and therefore $\tau_q = q-1$, as shown in Fig.~\ref{fig:mf} (d) in red. For a localized state instead, depicted in Fig.~\ref{fig:mf} for $V/t = 3.0$ (c), the probability distribution is mostly concentrated in one site where it takes the value one, while it is close to zero on all other sites. Then, $\tau_q$ will be zero for any value of $q$ ($q>1$), as we show in Fig.~\ref{fig:mf} (d) in green. Note that Fig.~\ref{fig:mf} (d) is computed for a fixed system size $N$ and does not take into account any scaling of $\tau_q$. 

When $D_q$ is constant for all values of $q$ but takes a value between $0$ and $1$, we say that the state is fractal and it has only one fractal dimension, i.e. the structure can be defined with only one scaling exponent. Nevertheless, one can also encounter states for which $D_q$ depends on the value of $q$. Such states are multifractal, which is a generalization of fractal states. This is the case of the ground state depicted in Fig.~\ref{fig:mf}(b) for $V/t = 1.0$. In Fig.~\ref{fig:mf} (d) in blue, we see that now $\tau_q$ has a dependence on $q$ that is not linear. In Fig.~\ref{fig:mf}(e), we show the value of the $q$-fractal dimensions for different values of $q$, which decreases as we increase the value of the moments. The set of $q$-fractal dimensions completely characterizes the multifractal nature of the state. In particular, as we will see in the following Section, the fractal dimension $D_2$ is the most common one to use, since it gives us information on how the probability distribution of a given state fills the physical support of the system.

Another fundamental measure of multifractality which is more compact than the $q$-fractal dimensions is the multifractal spectrum $f(\alpha)$. Specifically, the number of sites $n$ where the onsite occupation fulfils $p_{n,j} \sim N^{-\alpha}$ scales as $N^{f(\alpha)}$. This quantity is interesting because it contains all the information about $\tau_q$ via the Legendre transformation. In order to compute it numerically, we compute numerically the gradient of $\tau_q$ for fixed integer values of $q^*$ such that
 \begin{equation}
     \alpha = \frac{d\tau_q}{dq}\Big\vert_{q^*}.
 \end{equation}
Then, $\alpha_\text{max}$ will be the maximum slope of the function $\tau_q (q)$ for the chosen discretization. Next, we compute $f(\alpha) = q^*\alpha - \tau_{q^*}$~\cite{roy2018}. Figure.~\ref{fig:mf} (f) depicts the multifractal spectrum ($f(\alpha)$ in terms of $\alpha$) for the ergodic, localized and multifractal states. For this computation, we have used $q\in [-10,10]$. Here it is important to note that the multifractal spectrum is affected by finite size effects and that one can compute the extrapolation to the thermodynamic limit as in~\cite{deluca2014,Deng2016}, but even for finite systems as the one under consideration, we can already see that the three states have very different multifractal spectrums. For ergodic states (red), the multifractal spectrum will be concentrated around one, converging to a delta function at $1$ for $N\rightarrow \infty$. For localized states instead, $f(\alpha) = \lim_{\alpha_\text{max} \rightarrow \infty} \alpha/\alpha_\text{max}$, which leads to a linear dependence for small values of $\alpha$. Most importantly, the multifractal spectrum is parabolic for multifractal states, with a maximum value $f(\alpha) = 1$. The spectrum for multifractal states is also known to satisfy the relation $f(1+\alpha) = f(1-\alpha) + \alpha$.

\subsubsection{Mean fractal dimension $\bar{D}_2$}

\begin{figure}[ht]
\includegraphics[width=0.8\columnwidth]{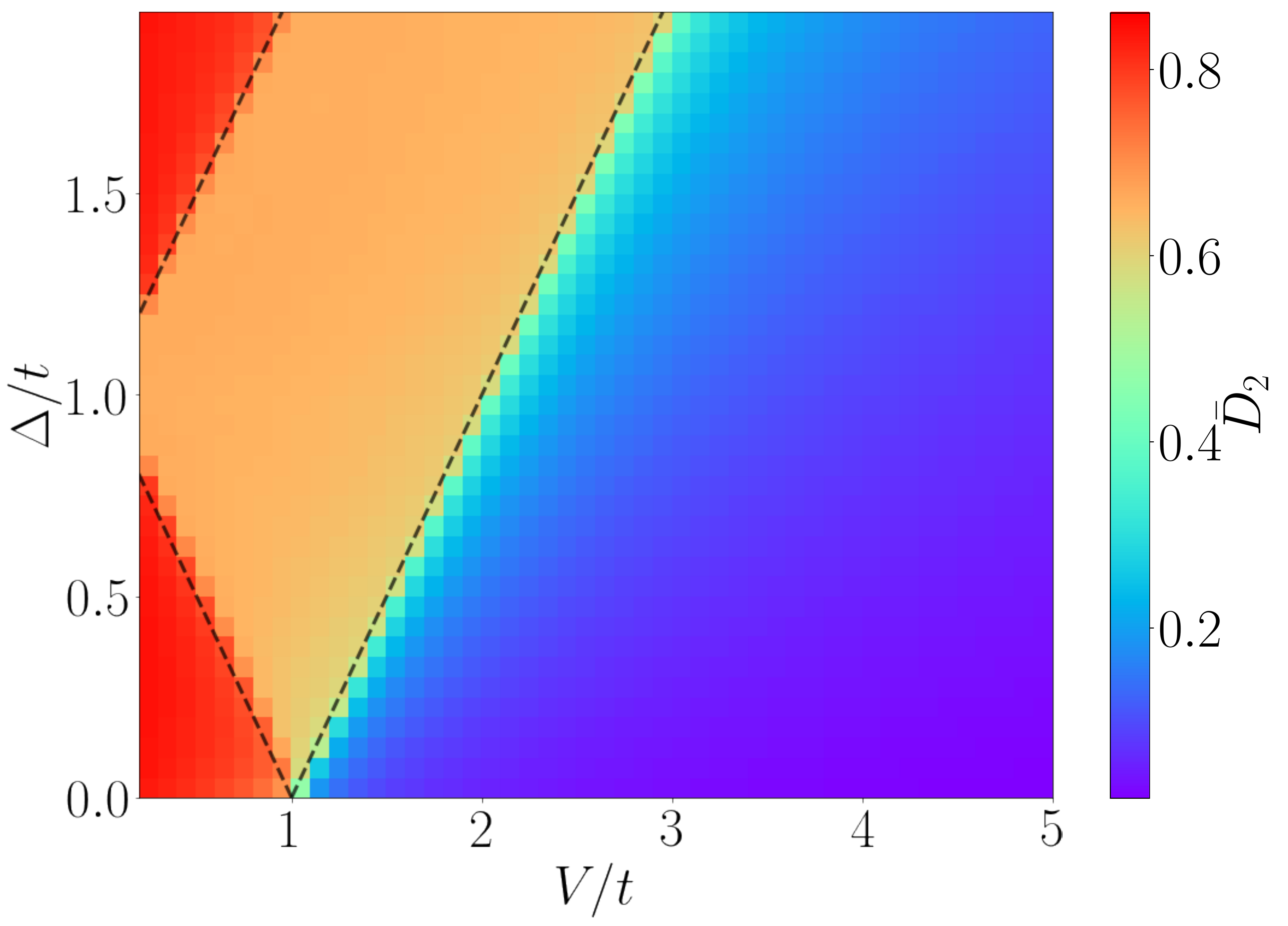}
\caption{\label{fig:gmeasures_D2} \textbf{Mean fractal dimension $D_2$:} Phase diagram for the Hamiltonian in Eq.~\eqref{eq: hamiltonian_SR_AAH}, which corresponds to the limit $\xi, \alpha \rightarrow \infty$,  using the mean fractal dimension $\bar{D}_2$ for different values of $V/t$ and $\Delta/t$ and fixing $N = 1597$. We confirm that the multifractal intermediate region is comprised between $V_t = \vert \Delta \pm t\vert$ (dashed lines).}
\end{figure}

Going back to the problem of distinguishing between ergodic, localized and multifractal eigenstates, let us now focus on the $2$-fractal dimension $\bar{D}_2$, which is an alternative to the participation ratios that does not require a finite size extrapolation to classify the different types of states. As we explained above, the most common fractal dimension is $D_2$, which is equivalent to the usual box-counting fractal dimension~\cite{nilanjan21} and it gives us information on how the probability distribution of a state fills the space. In particular, it can be related to the participation ratios as
\begin{equation}\label{eq: D2}
    D_2(n) = -\frac{\log{\text{IPR}_n}}{\log{N}},
\end{equation}
where $\text{IPR}_n\equiv\sum_{j=1}^{N} (\vert u_{n,j}\vert^{2} + \vert v_{n,j}\vert ^{2})^2$ corresponds to the IPR of a given state $n$. Here we are interested in using the mean fractal dimension $\bar{D}_2$ over all eigenstates, which reads
\begin{equation}
    \bar{D}_2 = -\frac{1}{2N} \sum_{n=1}^{2N} \frac{\log{\text{IPR}_n}}{\log{N}}. 
\end{equation}
The fractal dimension $D_2$ is the most used quantity in the literature to distinguish between localized, ergodic and multifractal states \cite{deng2019, nilanjan21, xu21}. Without the need to extrapolate its value to the thermodynamic limit, it is expected to be $1$ for a given localized state, $0$ for an extended state and in between zero and one for a multifractal state. Figure~\ref{fig:gmeasures_D2} shows the phase diagram of the Hamiltonian $H_{\xi, \alpha \rightarrow \infty}$ in terms of the superconducting pairing $\Delta$ and the quasi-periodic potential strength $V$. As mentioned above, the transitions are found on the lines $V_t = \pm \vert\Delta-t\vert$ (black dashed lines).

\subsection{Local characterization}\label{sub:lmethods}

In this Section, we introduce local methods to characterize models, whose spectra show energy-dependent localization properties. In particular, they allow us to characterize the existence of mobility edges or mixing of bands with different localization properties. In particular, we consider the fractal dimension $D_2$ and the energy level spacings. Finally, we also comment on possible theoretical approaches to describe energy-dependent metal-insulator transitions.

\subsubsection{Long range hopping limit}

We illustrate these methods by applying them to the Hamiltonian~\eqref{eq: hamiltonian} within the limit of $\Delta = 0$
\begin{equation}\label{eq: hamiltonian_PL_AAH}
    H_{\Delta = 0} = -\sum_{i=0}^{N-1}\left(\sum_{l=1}^{N-1}\frac{t}{l^\xi}c^\dagger_{i+l}c_{i} + h.c.\right) + 
    2V f(i)c^\dagger_{i}c_{i}.
\end{equation}
We consider periodic boundary conditions (PBC) and solve the eigenvalue problem using exact diagonalization for different values of the on-site quasi-periodic potential $V$. Note that for this case we do not use the Bogoliubov-de-Gennes (BdG) basis. This limit is equivalent to the Aubry Andr\'e model with power law hoppings studied in Refs.~\cite{deng2019, nilanjan21}. This model is known to present a mobility edge which, for a fixed value of the quasi-periodic potential $V/t$, separates ergodic-localized states when $\xi \geq 1$ or ergodic-multifractal states for $\xi \leq 1$. Moreover, the mobility edge for the Hamiltonian~\eqref{eq: hamiltonian_PL_AAH} can be described in terms of different regimes: for a quasi-periodic potential $V/t$ smaller than a critical value $V_1/t$, the transition between ergodic-localized or ergodic-multifractal states happens at a fixed value of the filling $n/N = \beta$. Here $n$ is the number of states below the mobility edge, $N$ is the total number of states or the length of the system, and $\beta$ is the modulation frequency in Eq.~\eqref{eq:modulation}. Then, for a quasi-periodic potential in between $V_1/t$ and a second critical value $V_2/t$, the transition happens at a fixed value of the filling $n/N = \beta^2$, and we successively define different regions for  $n/N = \beta^s$, $s\in \mathbb{N}$. Moreover, each of this fillings corresponds to a gap between sets of energy bands. 

\subsubsection{Fractal dimension $D_2$}

\begin{figure}[ht]
\includegraphics[width=\columnwidth]{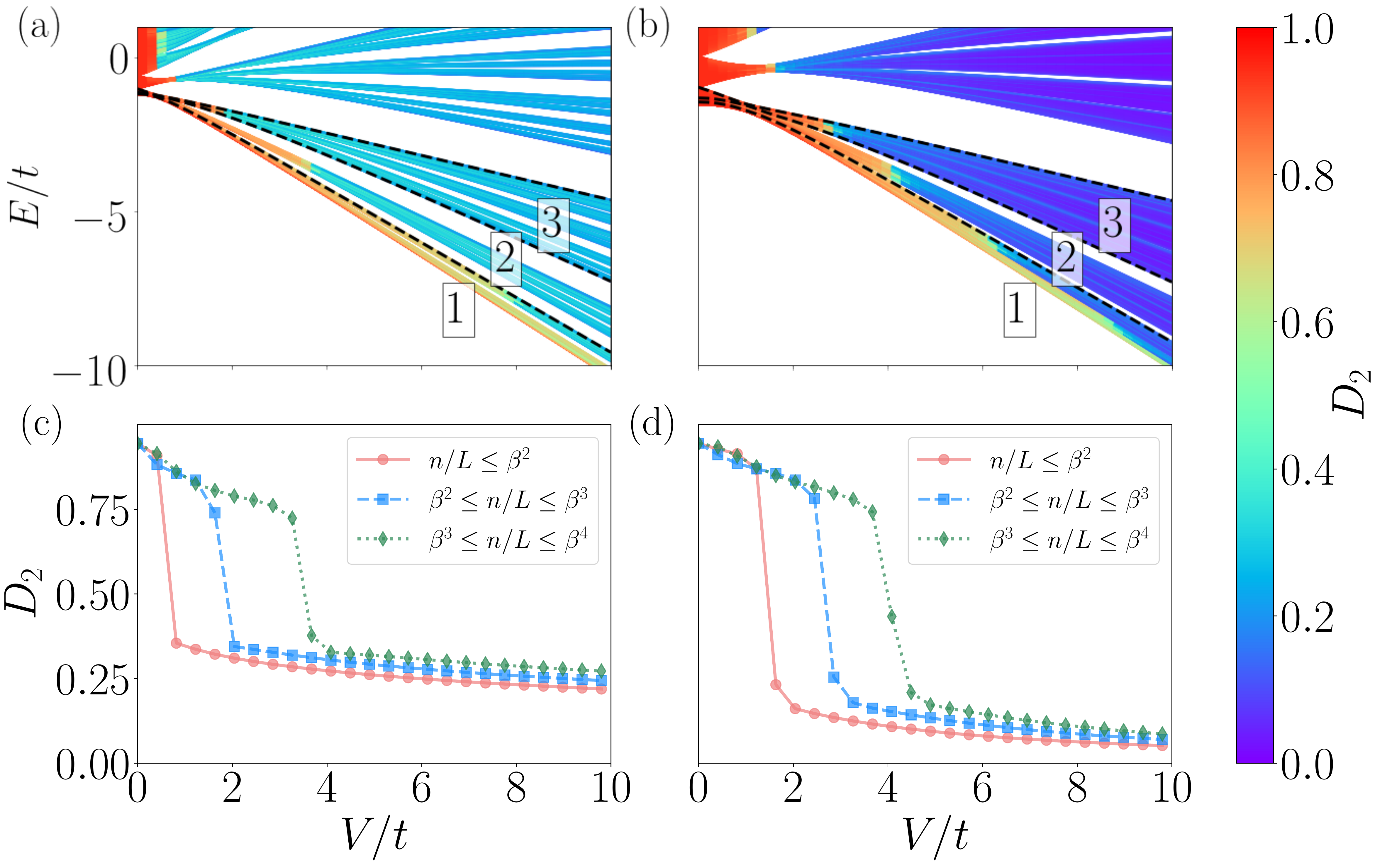}
\caption{\label{fig:lmeasures} \textbf{Study of a mobility edge:} \textbf{(a-b)} Energy spectrum for the Hamiltonian in eq.~\eqref{eq: hamiltonian_PL_AAH} for $N = 1597$ and $\xi = 0.5$ (a) or  $\xi = 1.5$ (b) in terms of the quasi-periodic potential $V/t$. The color map indicates to the fractal dimension $D_2$ of each eigenstate corresponding to the energy $E/t$. We show an energy-dependent transition between ergodic-multifractal states (a) and ergodic-localized states (b). The black dashed lines separating the bands indicate the fillings $n/N = \beta^2$ (1), $n/N = \beta^3$ (2) and $n/N = \beta^4$ (3), where $\beta$ is the modulation frequency from eq.~\eqref{eq:modulation}. \textbf{(c-d)} Mean fractal dimension $\bar{D}_2$ for the states below $n/N = \beta^2$, between $n/N = \beta^2$ and $n/N = \beta^3$ and between $n/N = \beta^3$ and $n/N = \beta^4$, for $\xi = 0.5$ (c) and $\xi = 1.5$ (d). For each band, there is a sharp transition at a different critical value of the quasi-periodic potential $V/t$.}
\end{figure}

To study the model described above, we are generally interested in distinguishing between different bands. Therefore, one has to compute quantities specific to a band. In particular, we can still rely on the fractal dimension $D_2$, but instead of taking a mean value of this quantity we study it for each eigenstate $n$
as in Eq.~\eqref{eq: D2}.
Figures~\ref{fig:lmeasures}(a,b) show the energy spectrum for $\xi = 0.5$ (a) and for $\xi = 1.5$ (b), where the color map indicates the fractal dimension $D_2$ of the state corresponding to each energy $E/t$. Figure~\ref{fig:lmeasures}(a) shows a clear mobility edge separating ergodic states ($D_2 = 1$) from multifractal states ($D_2 \sim 0.3$), as indicated in~\cite{deng2019, nilanjan21}. Figure~\ref{fig:lmeasures}(b) depicts a mobility edge separating ergodic states from localized states ($D_2 = 0$). In order to study the block-like structure of the mobility edge, we study the different bands separately. In particular, the black dashed lines in Figs.~\ref{fig:lmeasures}(a-b) separate different bands and correspond to $n/N = \beta^s$ for $s \in [2, 3, 4]$, where $n/N$ is the filling and $\beta$ is the modulation frequency from Eq.~\eqref{eq:modulation}. Figures~\ref{fig:lmeasures}(c-d) show the mean fractal dimension $\bar{D}_2$ for each set of energy bands $1$, $2$ and $3$ for both $\xi = 0.5$ (a) and $\xi = 1.5$ (b). We clearly see that the transition happens at a different value of the quasi-periodic potential $V/t$ for each band. The transition is sharp, indicating that all the states in one band become localized or multifractal at the same time.

\subsubsection{Energy level spacing}

\begin{figure}[ht]
\includegraphics[width=\columnwidth]{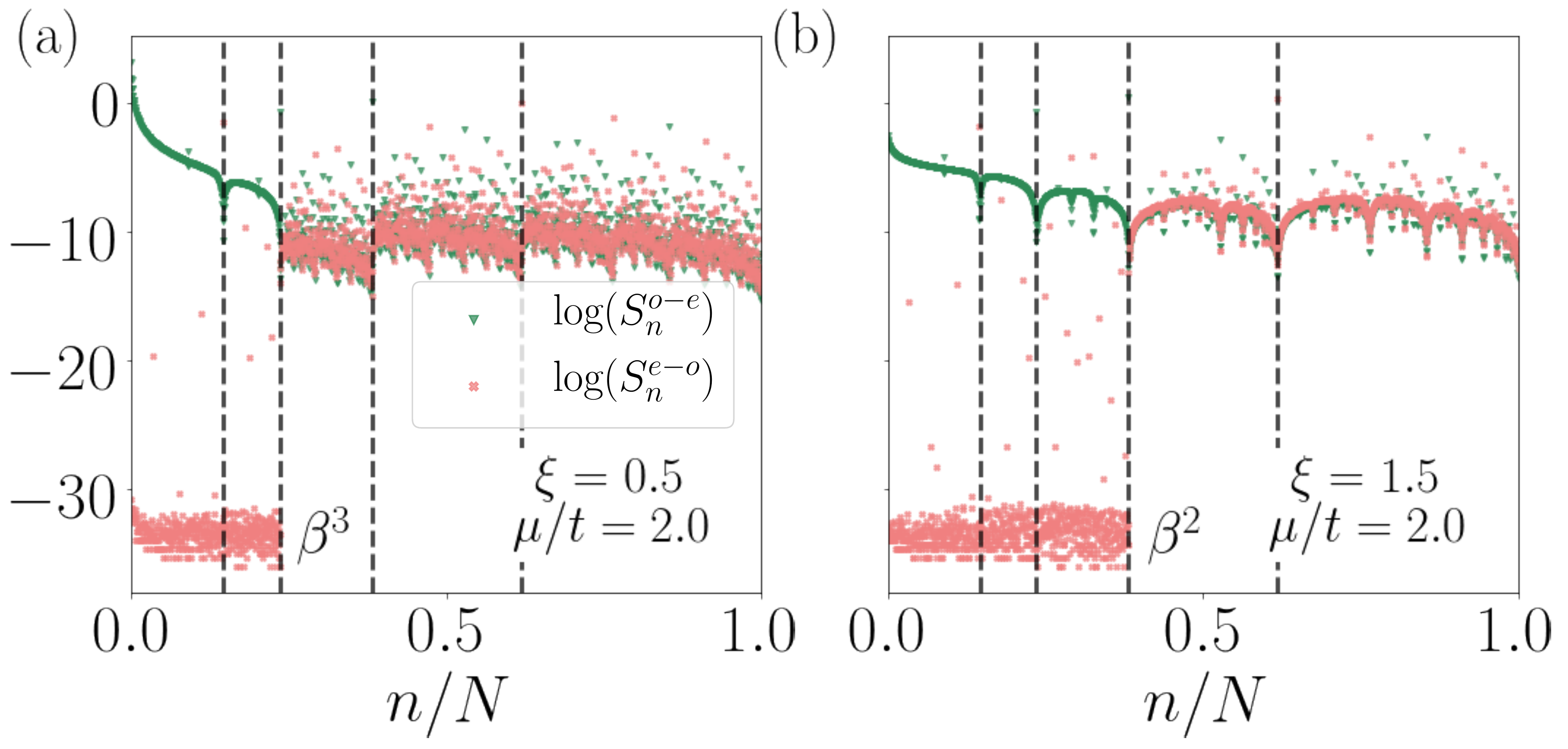}
\caption{\label{fig:els_delta0} \textbf{Study of the energy level spacings:} \textbf{(a-b)} Logarithm of the even-odd(odd-even) energy level spacings in green(red) for the Hamiltonian~\eqref{eq: hamiltonian_PL_AAH} with $\xi = 0.5$ and $V/t = 1.5$ (a), with a multifractal-ergodic mobility edge at $n/N = \beta^3$, and with $\xi = 1.5$ and $V/t = 1.0$ (b), with a localized-ergodic mobility edge at $n/N = \beta^2$. We fix $N=4181$.}
\end{figure}

Finally, in order to distinguish between localized, ergodic and multifractal regions of the spectrum one can also compute the energy level spacings as proposed in Ref.~\cite{deng2019}. Contrary to the methods indicated above, the study of the energy level spacings relies exclusively on the distribution of the eigenenergies $E_n$, instead of relying on the structure of the eigenstates. Therefore, it is an interesting quantity to complement our analysis. In particular, we compute the even-odd (odd-even) spacings
\begin{eqnarray}\label{eq:els}
    S_n^{e-o} &&= E_{2n}-E_{2n-1}, \nonumber \\ 
    S_n^{o-e} &&= E_{2n+1}-E_{2n},
\end{eqnarray}
where $E_n$ are the energy levels for  $n={1,...,N}$. In the Bogoliubov-de-Gennes(BdG) basis for $\Delta \neq 0$, we take only the positive energies $E>0$, and then $E_1$ is the minimum positive energy. For ergodic regions of the energy spectrum, we expect the eigenenergies to be doubly degenerate. Thus, $S_n^{e-o} = 0$ while $S_n^{o-e} \neq 0$ and we expect to see a gap between the the two. This can be seen in Fig.~\ref{fig:els_delta0} for the Hamiltonian~\eqref{eq: hamiltonian_PL_AAH}. The ergodic regions of the spectrum are located at $n/N < \beta^3$ for $\xi = 0.5$ and $V/t = 1.5$ (a) and at $n/N < \beta^2$ for $\xi = 1.5$ and $V/t = 1.0$ (b). For localized states this is not the case, and thus we will expect to see no difference between the even-odd(odd-even) cases. Again, this can be pictured in Fig.~\ref{fig:els_delta0}(b) for $n/N > \beta^3$.  Finally, multifractal regions of the spectrum will show strongly scattered distributions of both the odd-even and the even-odd cases, as seen in Fig.~\ref{fig:els_delta0}(a) for $n/N > \beta^3$. The full characterization of this model using the energy level spacings can be found in Ref.~\cite{deng2019}. This method allows us to study the whole energy spectrum with no previous knowledge of the different regions that might be coexisting, and the results coincide with the information obtained from studying the fractal dimension $D_2$.

\subsubsection{Generalizations of the self-duality conditions}

Most of the theoretical approaches on metal-insulator transitions in incommensurate systems rely on the notion of self-duality conditions. For the Aubry-Andr\'e model, the transition point between ergodic and localized states happens at a self-dual point, meaning that the Hamiltonian and its Fourier transform can be mapped to each other. Then, the eigenstates of the Hamiltonian can neither be localized nor extended, and one finds that critical behavior leads to multifractal states. Previous works have tried to generalize the condition of self-duality to more complex systems with energy-dependent metal-insulator transitions with relative success: 

\begin{itemize}
    \item For systems with superconducting pairing, in particular for the short-range Kitaev chain, self-duality conditions can only be found only for special points of the phase diagram in Fig.~\ref{fig:gmeasures_D2}, corresponding to $V/t = 0$ and $\Delta/t = 1$ and $V/t=1$ and $\Delta/t =0$ ~\cite{wang2016}. Nevertheless, these conditions have not been extended to describe the whole multifractal regime, for which an understanding of the origin of the critical properties is still not clear. 
    \item For systems with long-range hopping, self-duality conditions can be generalized in a few specific cases corresponding to only next nearest neighbor hoppings, Gaussian or power-law decaying hoppings~\cite{biddle2011}. In particular, for power-law decaying hoppings the slope can be approximated by the relation $E = V \cosh{p}-e^{p}$ where $p = \xi \ln{2}$, but this approximation becomes unreliable when the decay rate with respect to the distance is slow (usually $\xi<2$). 
    \item In some cases~\cite{Deng2016}, one can also take periodic approximants of a long-range quasi-periodic system in order to study the energy dispersion of each of the subbands in momentum space. By doing this, one can see that, with the right renormalization of the hopping and the decay exponent amplitude, each subband of the model can be effectively mapped to an AAH model, for which the existence of a metal-insulator transition can be explained in terms of self-duality. Again, this reasoning seems to be valid only when the decay exponent is sufficiently large. 
    \item  Alternatively, and specifically for power-law systems, there exist other theoretical approaches to get an insight into these kinds of transitions. One example, following~\cite{nilanjan21}, is to describe the structure of the phase diagram through the mathematical properties of the modulation frequency, but this is not easily generalized to more complex Hamiltonians. 
\end{itemize}

Even though some of the approaches we list above can be used get an insight of specific limits of the model under consideration, we see that a general theoretical framework for the study of these type of transitions is still missing, and therefore it is important to analyze and characterize models that go beyond the currently known examples. This is the case of the Hamiltonian in eq.~\eqref{eq: hamiltonian}, for which an exact solution for the self-duality conditions does not exist and one cannot rely on simple approximations. In the following Sections, we focus on the numerical analysis and characterization of novel properties which go beyond the already known limits.

\section{\label{sec:results}Results}

In this Section, we study the localization properties of the Hamiltonian~\eqref{eq: hamiltonian}. While previous studies~\cite{deng2019, nilanjan21, wang2016} focused on the limits of short-range superconducting pairing and long-range hopping amplitudes, we here study the effect of the long-range superconducting pairing term.  We therefore consider the limit $\xi \rightarrow \infty$ and study the localization properties of the model in terms of $V$, $\Delta$ and $\alpha$. In order to address the superconducting pairing term, we make use of the Bogoliubov-de-Gennes (BdG) basis. Moreover, to study the localization properties of the bulk we need to consider antiperiodic boundary conditions (APBC). The explicit form of the Hamiltonian with APBC can be found in Appendix~\ref{appendix:construction}.\\
We first characterize different regions of the phase diagram of our model globally. In particular, we use the mean fractal dimension $\bar{D}_2$ to characterize the energy spectrum as a whole, without considering energy-dependent transitions [similarly to the limit in Eq.~\eqref{eq: hamiltonian_PL_AAH}]. Then, we focus on the different sets of energy bands and study each eigenstate separately, in order to detect potential mobility edges or specific trends in the mixing of different types of localization in the energy spectrum. Finally, we further analyse the distribution of the fractal dimensions in the multifractal regions.

\subsection{Global characterization of the phase diagram of the system}

\begin{figure}[ht]
\includegraphics[width=\columnwidth]{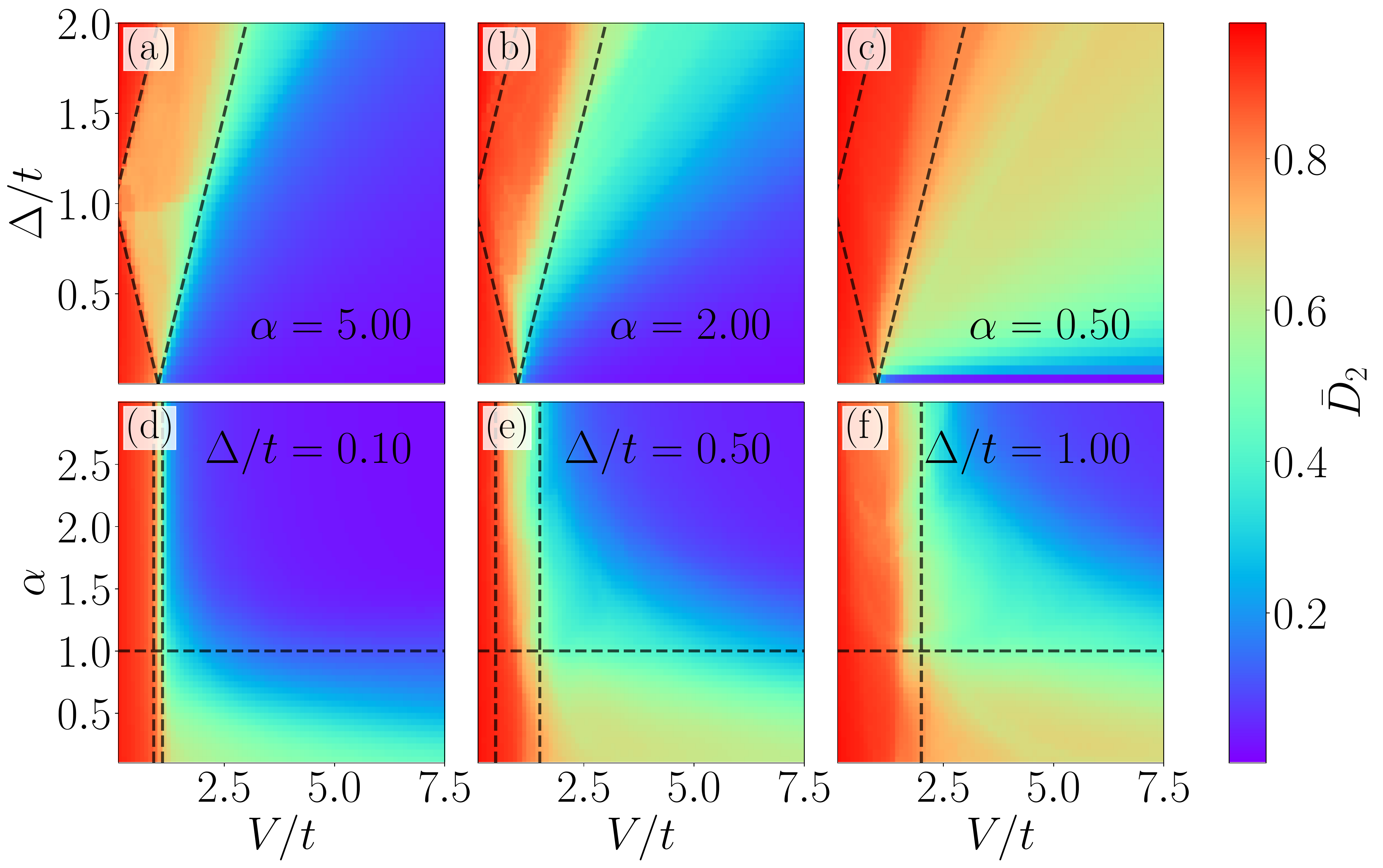}
\caption{\label{fig:phase_diagram} \textbf{Phase diagram of the model with $\xi \rightarrow \infty$:} \textbf{(a-c)} Mean fractal dimension $\bar{D}_2$ for $\alpha = 5.0$ (a), $\alpha=2.0$ (b) and $\alpha=0.5$ (c). The dashed lines at $\vert \Delta \pm t \vert$ enclose the multifractal intermediate region for the limit $\alpha, \xi \gg 1$.  For all the plots we consider the Hamiltonian in eq.~\eqref{eq: hamiltonian} with $N = 1597$ and $\xi \rightarrow \infty$. \textbf{(d-f)} Mean fractal dimension $\bar{D}_2$ for $\Delta/t = 0.1$ (d), $\Delta/t = 0.5$ (e) and $\Delta/t = 1.0$ (f). The horizontal dashed line at $\alpha = 1$ separates the regions with ergodic-multifractal and ergodic-localized transitions for the limit $\Delta/t \rightarrow \infty$, while the vertical dashed lines at $\vert \Delta \pm t \vert$ enclose the multifractal intermediate region for the limit $\alpha, \xi \gg 1$. }
\end{figure}

We first characterize the system globally with the help of the mean fractal dimension. We aim to obtain a general picture of the phase diagram of our model for different cuts in the parameter space, in order to identify regions that should be explored in more detail. Figures~\ref{fig:phase_diagram}(a-c) show the mean fractal dimension $\bar{D}_2$  for fixed values of the decay exponent, $\alpha = 5.0$ (d), $\alpha=2.0$ (e) and $\alpha=0.5$ (f) and in terms of the quasi-periodic potential $V/t$ and the superconducting pairing $\Delta/t$, while in Figures~\ref{fig:phase_diagram}(d-f) we study the phase diagram for three fixed values of the superconducting pairing, $\Delta/t = 0.1$ (d), $\Delta/t = 0.5$ (e) and $\Delta/t = 1.0$ (f) and in terms of the quasi-periodic potential $V/t$ and the decay exponent $\alpha$.

For large values of $\alpha$ [Fig.~\ref{fig:phase_diagram}(a)], we recover the phase diagram of Fig.~\ref{fig:gmeasures_D2}, where the multifractal region is delimited by $\vert \Delta \pm t\vert$ (see black dashed lines), as expected. Nevertheless, in contrast with the limiting case, here the boundaries between the multifractal region (in orange) and the localized region (in blue) are not sharp. This indicates that the transitions are energy-dependent and therefore not all the eigenstates of the system have the same localization properties, which we will analyse further in the next section. In particular, this behavior is enhanced when decreasing $\alpha$ [Fig.~\ref{fig:phase_diagram}(b)], to the point that the multifractal region seen in orange in Fig.~\ref{fig:phase_diagram}(a) can be barely identified in Fig.~\ref{fig:phase_diagram}(b). Moreover, for $\alpha < 1$ the system is never localized [Fig.~\ref{fig:phase_diagram}(c)], with the exception of the line corresponding to $\Delta/t = 0$, where we recover the AAH model. The transition between ergodic and multifractal states for a decay exponent $\alpha$ smaller than one is similar to the results shown in Fig.~\ref{fig:lmeasures}(a), which correspond to the Hamiltonian $H_{\Delta = 0}$ and the case $\xi < 1$. We can therefore conclude that the presence of long-range, either in the hopping or the superconducting pairing, with a decay exponent $\alpha\ll1$, induces an ergodic to multifractal transition of the energy spectrum when increasing the quasi-periodic potential $V/t$. This is confirmed in Figures~\ref{fig:phase_diagram}(d-f), for $\alpha < 1$ (below the dashed horizontal line), the system undergoes a transition from ergodic to multifractal states for any finite value of $\Delta$, while for $\alpha > 1$ there is a transition from ergodic to localized states with an intermediate multifractal region in between. The vertical dashed lines enclose the region between $\vert \Delta \pm t\vert$. Furthermore, we notice that for very small values of $\Delta$ [Fig.~\ref{fig:phase_diagram}(d)], the multifractal region starts to vanish. The latter is expected for the AAH limit. Furthermore, the boundaries between the different regions are not sharp, which again indicates that the transition point might be different for different eigenstates. In particular, this energy dependence is enhanced for larger values of $\Delta$, as seen in Fig.~\ref{fig:phase_diagram}(f), and prevents the study of the multifractal intermediate regions using global characterization methods. In the following sections, we will use the local characterization methods from Section.~\ref{sub:lmethods} to characterize individually the eigenstates and study the transition points in between regions in more detail.

\begin{figure*}
\centering
\includegraphics[width=0.65\textwidth]{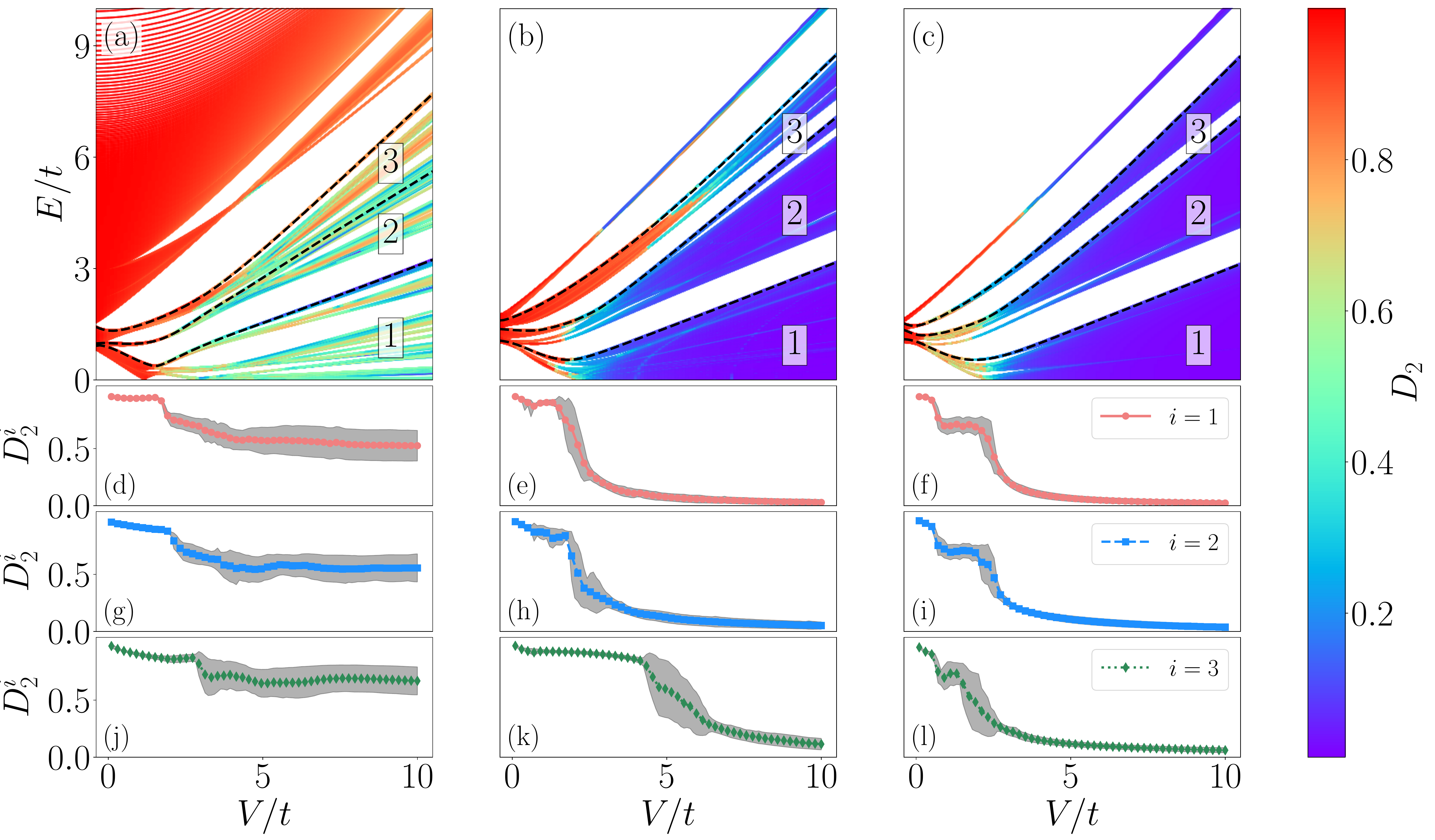}
\caption{\label{fig:bands}\textbf{Study of the energy-dependent transitions:} \textbf{(a-c)} Positive energy spectrum of the Hamiltonian in eq.~\eqref{eq: hamiltonian} for different values of the chemical potenital $V/t$. The color indicates the fractal dimension $D_2$ of the eigenstate that corresponds to each energy $E/t$, computed for $\alpha = 0.5$ (a), $\alpha=2.0$ (b) and $\alpha=5.0$ (c). The dashed lines separate four different sets of energy bands for lower energy states (1), intermediate (2) and high energy states (3). \textbf{(d-l)} Mean fractal dimension $\bar{D}_2^i$ for the states in the first ($i=1$), second ($i=2$) and third ($i=3$) set of bands and for $\alpha = 0.5$ (d,g,j), $\alpha=2.0$ (e,h,k) and $\alpha=5.0$ (f,i,l). The grey shaded area indicates the standard deviation $\pm \sigma$. For all the plots we take $N = 4181$, $\Delta/t = 0.8$ and $\xi \rightarrow \infty$. }
\end{figure*}

\subsection{Local characterization of the energy-dependent transitions}

We now analyze more in depth the behavior of sets of energy bands separately in order to properly characterize the transitions between states with different localization properties. Figures~\ref{fig:bands}(a-c) depict the positive part of the energy spectrum in terms of the quasi-periodic potential $V/t$ for a fixed value of $\Delta=0.8t$ \footnote{Here we use the BdG basis to treat the non-vanishing superconducting pairing term. This leads to an energy spectrum which is particle-hole symmetric, which allows us to show only positive energies.} and three different values of the decay exponent $\alpha=0.5$ (a), $\alpha=2$ (b), $\alpha=5$ (c). Here, the color map indicates the fractal dimension $D_2$ of each eigenstate. \\
Even though the transition is, as expected, energy dependent in all three cases, we observe no clear mobility edge that can be approximated by either a line or a block-like structure such as the one in Fig.~\ref{fig:lmeasures} for the Hamiltonian  $H_{\Delta = 0}$. Instead, states with different fractal dimension can be observed in Fig.~\ref{fig:bands}(a). In Fig.~\ref{fig:bands}(b), the transition happens at different values of the quasi-periodic potential $V/t$ which do not necessarily increase when considering higher energy states. Moreover, having particle-hole symmetry prevents the system for a fixed $V/t$ from having only one transition for positive or negative energies. Indeed, in the BdG basis and for the limit $\Delta = 0$, we see a superposition of two mobility edges that can be approximated by lines with opposite slopes. Then, for finite values of $\Delta$ the two transitions hybridize, leading to mixing of different types of bands.\\
Figs.~\ref{fig:bands}(d-l) show the mean fractal dimension $\bar{D}_2^i$ within a set of energy bands together with the standard deviation $\sigma$ (shaded grey area), where the index $i=1,2,3$ indicates the different sets with increasing energy shown in Figs~\ref{fig:bands}(a-c). We consider distinct energy bands to be separated by an energy gap that does not decrease as we increase the system's size. In this case, the separation between the bands does not necessarily happen at a filling $n/N = \beta^s$ as it was reported for $\Delta = 0$~\cite{deng2019}. In Fig.~\ref{fig:bands}(a-c) we separate between sets of energy bands where we find the largest gap between adjacent eigenenergies at $V/t=2$. \\
Analyzing transitions at different sets of energy bands, we can conclude the following. First, we confirm that for  $\alpha = 0.5$ [panels (d,g,j) of Fig.~\ref{fig:bands}] the system undergoes a transition from ergodic to multifractal states. In particular, the multifractal states show different fractal dimensions ranging from $\bar{D}_2 = 0.5$ to $\bar{D}_2 = 0.8$, which is confirmed by the large value of the standard deviation. Second, for $\alpha=2.0$ [panels (e,h,k) of Fig.~\ref{fig:bands}] and $\alpha=5.0$ [panels (f,i,l) of Fig.~\ref{fig:bands}], we observe a transition from ergodic to localized states with an intermediate multifractal region, whose size varies depending on the energy of the band. Again, the fractal dimension of the intermediate region shows a high standard deviation, which is negligible in the ergodic and the localized regions.  Third, for all values of $\alpha$, the general trend is that the transition moves towards larger values of the quasi-periodic potential $V$ as we consider higher energies, similarly to what was reported for $\Delta = 0$ in Fig.~\ref{fig:lmeasures}(c-d). \\
The superconducting pairing therefore leads to the existence of an intermediate multifractal region, which becomes energy dependent when the pairing is long-range. As a consequence, we see hybridization of bands with different types of localization properties, similar to Ref.~\cite{rossignolo19}. In particular, this means that the system described by the Hamiltonian in eq.~\eqref{eq: hamiltonian} can host coexisting localized, multifractal and ergodic states, which was not possible in the limits of the model that have been previously studied \cite{deng2019, nilanjan21, yahyavi19, zeng16, lv2022}. Such regimes can be understood as the system being in a mix of regimes in energy space, where different bands show different localization properties. In the following paragraphs, we zoom into a specific case with fixed $V/t$ in order to study the distribution of the fractal dimensions $D_2$ and to further investigate whether all types of localization can coexist.

\subsection{Further analysis of the effects of long-range superconducting paring}

Finally, we focus on the analysis of the effects arising from to the power-law superconducting pairing, which include the hybridization of energy bands with different types of localization that coexist for a system with fixed quasi-periodic potential $V/t$. 

\begin{figure}[t]
\includegraphics[width=\columnwidth]{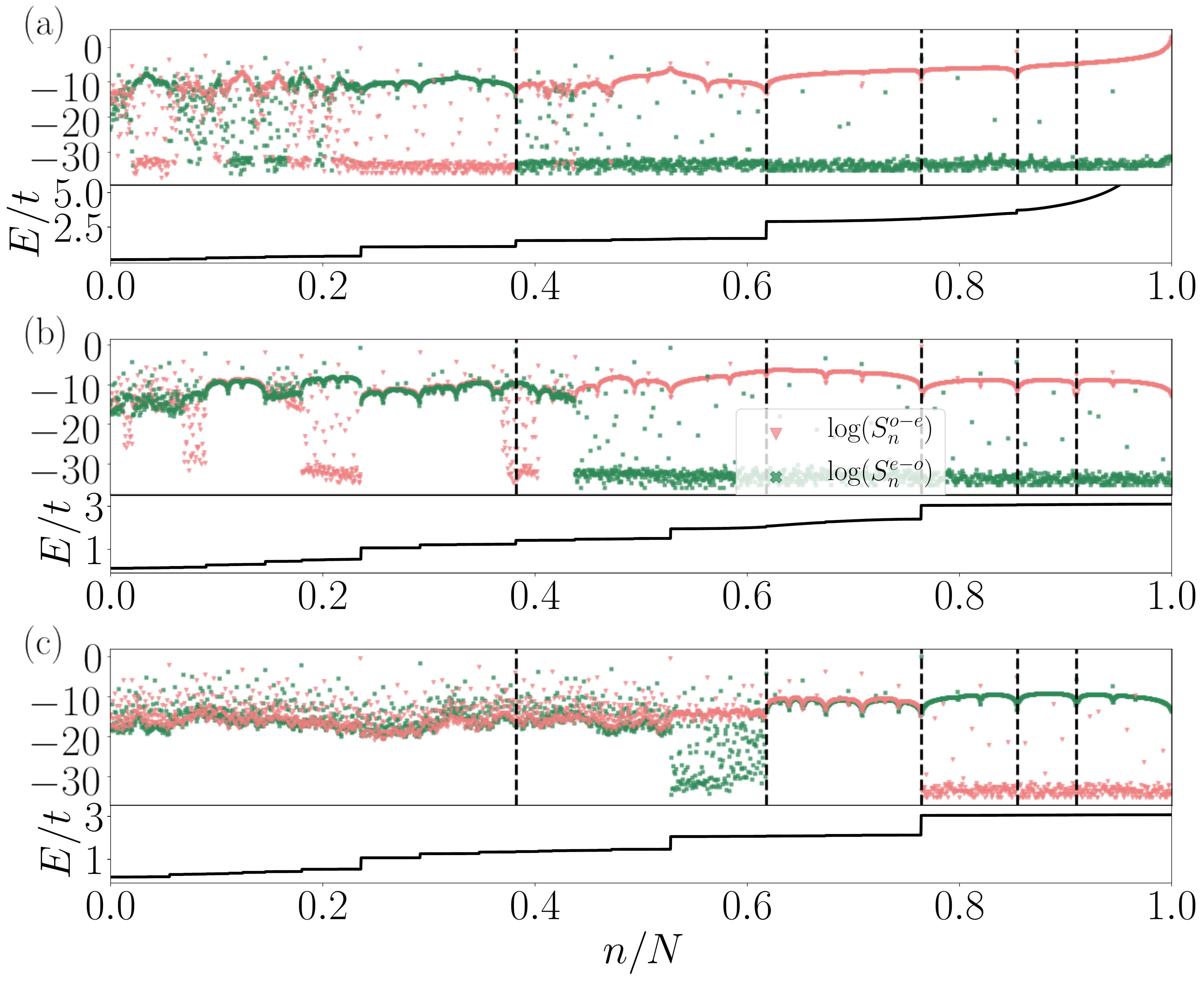}
\caption{\label{fig:els} \textbf{Energy level spacing:} \textbf{(a-c)} Logarithm of even-odd $S_n^{e-o}$ (odd-even $S_n^{o-e}$) energy level spacings in red(green) for the Hamiltonian in eq.~\eqref{eq: hamiltonian} and eigenenergies $E/t$ in black. The black dashed lines indicate powers of the modulation frequency $1-\beta^s$, for $s = 1,2,3,4,5$. We take $N = 4181$, $\Delta = 0.8$ and $\xi \rightarrow \infty$. We consider $\alpha=0.5, V/t = 2.0$ in (a) and $\alpha=2.0, V/t = 2.0$ in (b) and $\alpha=5.0, V/t = 2.0$ in (c). Some of the transitions coincide with values of $n/N = 1-\beta^s$ while others coincide with a jump in the eigenenergies, $E/t$, which indicates a gap between bands. In (a) we observe a transition between ergodic states to a region which is a mix of multifractal and ergodic bands, while in (b) and (c) we see a mix of ergodic, localized and multifractal bands.}
\end{figure}

\begin{figure}[ht]
\includegraphics[width=\columnwidth]{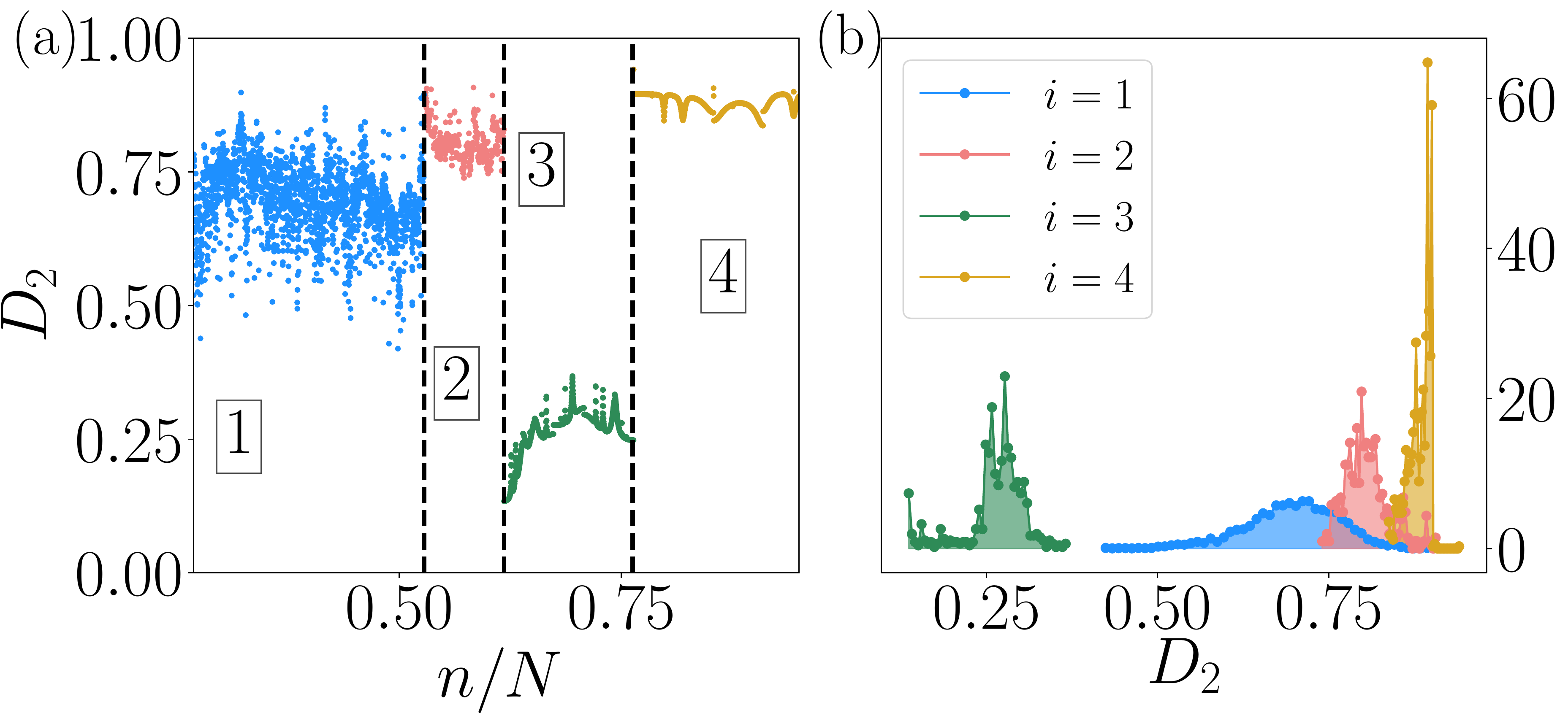}
\caption{\label{fig:histogram} \textbf{Distribution of fractal dimensions $D_2$:} \textbf{(a)} Fractal dimension $D_2$ for all the eigenstates with positive energies $E/t$ of the Hamiltonian in eq.~\eqref{eq: hamiltonian} for $\Delta = 0.8$  $\alpha = 5.0$, $V/t = 2.0$ and $\xi \rightarrow \infty$, indexed by the fraction $n/N$, where $n = 1,...,N$ and $N = 6765$ is the size of the system. There are four different regions corresponding to multifractal ($1$ and $2$), localized ($3$) and ergodic ($4$) states. The black dashed lines indicate the separation between the regions. \textbf{(b)} Histogram of the fractal dimensions $D_2$ for each of the regimes in (a), $i = 1,2,3,4$. In particular, the fractal dimensions for the multifractal regime ($1$) in blue are distributed around $D_2 = 0.7$, while regime ($2$) is peaked at $D_2 = 0.8$. }
\end{figure}

We first study the energy level spacing to identify the different regimes in the energy spectrum. In Fig.~\ref{fig:els}(a-c), we fix the value of the quasi-periodic potential at $V/t = 2.0$ and we study the positive energies of the system for $\Delta/t=0.8$ and $\alpha = 0.5$ (a), $\alpha=2.0$ (b), $\alpha=5.0$ (c). In particular, we are interested in using the quantitites defined in Eq.~\eqref{eq:els} to study whether there are regimes of band hybridization of ergodic, localized and multifractal states, and to characterize the exact location of the transitions. The black dashed lines correspond to different powers of the modulation frequency $1-\beta^s$, for $s = 1,2,3,4,5$. Here, we use $1-\beta^s$ instead of $\beta^s$ because the system shows particle-hole symmetry and $n/N=0$ corresponds to the energy $E/t = 0$. Below the energy level spacings, we plot the distribution of eigenenergies  $E/t$, in order to identify a change of energy band. Note that a jump in the eigenenergies can also be identified with an isolated point either in $S_n^{o-e}$ or $S_n^{e-o}$ which is larger than the neighboring level spacings, and thus coincides with an energy gap that marks the distinction between two energy bands. In Fig.~\ref{fig:els}(a) we see an ergodic to multifractal transition at $n/N \approx 0.2$. The transition coincides with a very small energy gap which is difficult to resolve. Nevertheless, we clearly observe a change of behavior in the odd-even (even-odd) spacings. For $n/N \lessapprox 0.2$ we see hybridization of ergodic and multifractal bands, usually separated by small energy gaps which correspond to a jump in the energies $E/t$. We also see that at $n/N = 0.38$, which corresponds to $n/N = 1-\beta$ and also to a clear energy gap, the roles of $S_n^{e-o}$ and $S_n^{o-e}$ are exchanged. This is an effect of the BdG basis, for which two copies of the energies of the system with different sign would hybridize and thus the counting of odd-even (even-odd) energy level spacings switches for certain values of the filling $n/N$. In Fig.~\ref{fig:els}(b), we report a transition from an ergodic to localized spectrum at $n/N \approx 0.45$, which coincides with a small energy gap. For $n/N \lessapprox 0.45$ we find alternating multifractal and localized bands separated by non-vanishing energy gaps. Finally, in Fig.~\ref{fig:els}(c) we can see a clear transition from ergodic to localized and then to multifractal states coexisting in the same system. The transitions now correspond to $n/N = 1-\beta^2$ and $n/N = 1-\beta^3$ respectively. We also observe a region, between $n/N \approx 0.5$ and $n/N = 1-\beta^2$, which shows an intermediate behavior between ergodic and multifractal states. In the following paragraph, we study the distribution of fractal dimensions and compare this region to the multifractal regime for $n/N \lessapprox 0.5$. From the study of the energy level spacings, we conclude that most of the transitions between different types of localization coincide with non-vanishing energy gaps which separate two bands, although most of these energy gaps are very small due to the multifractal character of the energy spectrum. Moreover, some of them can still be located in terms of the modulation frequency $\beta$, while others cannot. The system for a given fixed value of the quasi-periodic potential $V/t$ shows hybridization between bands with different types of localization, which can be described as mix of different regimes in energy space. \\
Lastly, we study the distribution of the fractal dimension $D_2$ in the different regimes of the mixed energy spectrum for $\alpha = 5.0$, $V/t = 2.0$ and $\Delta/t=0.8$, which have already been identified in Fig.~\ref{fig:els} (c). In Fig.~\ref{fig:histogram} (a), we report the fractal dimension $D_2$ for the eigenstates of the system within the interval from $n/N = 0.25$ to $n/N = 1$, where there are four distinct regimes. Region $(1)$ corresponds to regular multifractal behavior, while region $(2)$ corresponds to a regime which, following the study of the energy level spacings in Fig.~\ref{fig:els} (c), shows an intermediate behavior between multifractal and ergodic. Region $(3)$ shows localized behavior and region $(4)$ shows ergodic behavior. In Appendix~\ref{appendix:scaling} we show the scaling of the mean fractal dimension $\bar{D}_2$ for each of the regions with different behavior. In Fig.~\ref{fig:histogram} (b), we plot a histogram of the fractal dimension $D_2$ of the eigenstates of each region, in order to study how this quantity is distributed. The fractal dimensions of the ergodic regime, in yellow, are highly peaked around $1$, as expected. On the other hand, the localized regime, in green, is highly peaked around $D_2 = 0.25$, nevertheless we would expect to see a distribution around $D_2 = 0$. Indeed, this is what we observe for large values of $V/t$, but for the region of interest around $V/t = 2.0$, where we find hybridization of different bands, the eigenstates are still not fully localized. Nevertheless, the localized region is easily distinguished from the other regimes and it is also confirmed by the results in Fig.~\ref{fig:els} (c). We now focus on the multifractal regimes in blue and red. The multifractal regime in blue shows wide distribution of fractal dimensions centered around $D_2 = 0.7$, in contrast to results which were previously reported for systems with no superconducting pairing term~\cite{deng2019}, where the fractal dimension in the multifractal regime is highly peaked around one value. The region in red shows instead a different distribution which is peaked around $D_2 = 0.8$, with eigenstates that are not ergodic but extend further through the lattice. We have therefore characterized the multifractal regime of the system with power-law superconducting pairing, which stands out for displaying a wide distribution of fractal dimensions instead of a peak around a specific value. We also identify a novel multifractal regime, which is characterized with a different distribution of fractal dimensions and shows a distinct behavior in the study of the energy level spacings.

\section{\label{sec:conclusions} Conclusions and outlook}

We presented an in-depth study of the localization properties of the AAH model with power-law decaying hopping and superconducting pairing. We first reviewed a toolbox for the characterization of localization properties on two known limits of the Hamiltonian, namely the short-range Kitaev with AAH modulation~\cite{deng2019, nilanjan21} and the AAH model with power-law hopping~\cite{yahyavi19, zeng16, lv2022}. In particular, we focused on global characterization methods for systems with no mobility edges and local characterization methods to investigate energy dependent transitions and hybridization of bands with different types of localization behavior.

We then used these tools to show that the localization properties of the AAH model become nontrivial under the combined effects of long-range and superconducting pairing. In particular, we reported energy-dependent transitions from ergodic to multifractal states for a decay exponent $\alpha$ smaller than one and energy-dependent transitions from ergodic to localized states with an intermediate multifractal region when the decay exponent $\alpha$ is larger than one. Most importantly, we showed that the size of the intermediate multifractal region depends not only on the value of the superconducting pairing term $\Delta$, but also on the energy band. The transitions cannot be described through a mobility edge, but instead we reported hybridization of bands with different type of localization behavior. In particular, this can lead to systems that show a mix of different regimes in energy space with hybridization of localized, multifractal and ergodic energy bands. Moreover, we showed that there are two types of multifractal regimes, where the distribution of the fractal dimensions either has a large standard deviation or is peaked around a higher value of the fractal dimension. In the last scenario, we showed that such states led to a distribution of the energy level spacings which is intermediate between ergodic and multifractal behavior, which suggests that the multifractal states in this regime become more extended throughout the system. 

In the literature one can find several examples of experimental realization of Aubry-Andr\'e models using cold atom simulation~\cite{billy08, lahini09, luschen18, modugno10, roati08, lohse16, nakajima16, an20}, as well as proposals to obtain superconducting pairing terms \cite{kraus12a, buhler14}. As an outlook, it would be interesting to study how to realize effective long-range Hamiltonians which lead to the effects that we reported. In this context, it would be important to think about the scale, i.e. which is the smallest $N$ needed to observe hybridization of bands and different types of multifractal regions. It would also be interesting to study the propagation of information and transport properties~\cite{luitz2020} in the cases which show band hybridization or in the multifractal region for the case $\alpha<1$, which is characterized by a distinct distribution of multifractal dimensions compared to systems that were previously studied. In addition, one could also study the scaling of the entanglement entropy for the multifractal states of the system \cite{tomasi2020} to better characterize the distribution of the fractal dimensions for certain regimes of the phase diagram and to generalize the renormalization analysis approach presented in~\cite{Deng2016} to study separately different regions of the energy spectrum of our model.

\paragraph*{Methods} The code that has been used to obtain all the data and figures for the paper can be found in \cite{code}.

\paragraph*{Acknowledgments} We thank P. Sierant and D. Rakshit for fruitful discussions. We acknowledge support from ERC AdG NOQIA, State Research Agency AEI (“Severo Ochoa” Center of Excellence CEX2019-000910-S) Plan National FIDEUA PID2019-106901GB-I00 project funded by MCIN/ AEI /10.13039/501100011033, FPI, QUANTERA MAQS PCI2019-111828-2 project funded by MCIN/AEI /10.13039/501100011033, Proyectos de I+D+I “Retos Colaboración” RTC2019-007196-7 project funded by MCIN/AEI /10.13039/501100011033, Fundaci\'o Privada Cellex, Fundaci\'o Mir-Puig, Generalitat de Catalunya (AGAUR Grant No. 2017 SGR 1341, CERCA program, QuantumCAT \ U16-011424, co-funded by ERDF Operational Program of Catalonia 2014-2020), EU Horizon 2020 FET-OPEN OPTOLogic (Grant No 899794), and the National Science Centre, Poland (Symfonia Grant No. 2016/20/W/ST4/00314), Marie Sk\l odowska-Curie grant STREDCH No 101029393, “La Caixa” Junior Leaders fellowships (ID100010434), and EU Horizon 2020 under Marie Sk\l odowska-Curie grant agreement No 847648 (LCF/BQ/PI19/11690013, LCF/BQ/PI20/11760031, LCF/BQ/PR20/11770012).
AD further acknowledges the financial support from a fellowship granted by la Caixa Foundation (ID 100010434, fellowship code LCF/BQ/PR20/11770012).
\appendix
\section{\label{appendix:construction}Construction of the Bogolibov-de-Gennes Hamiltonian with antiperiodic boundary conditions (APBC)}
\begin{figure}[t]
\includegraphics[width=0.7\columnwidth]{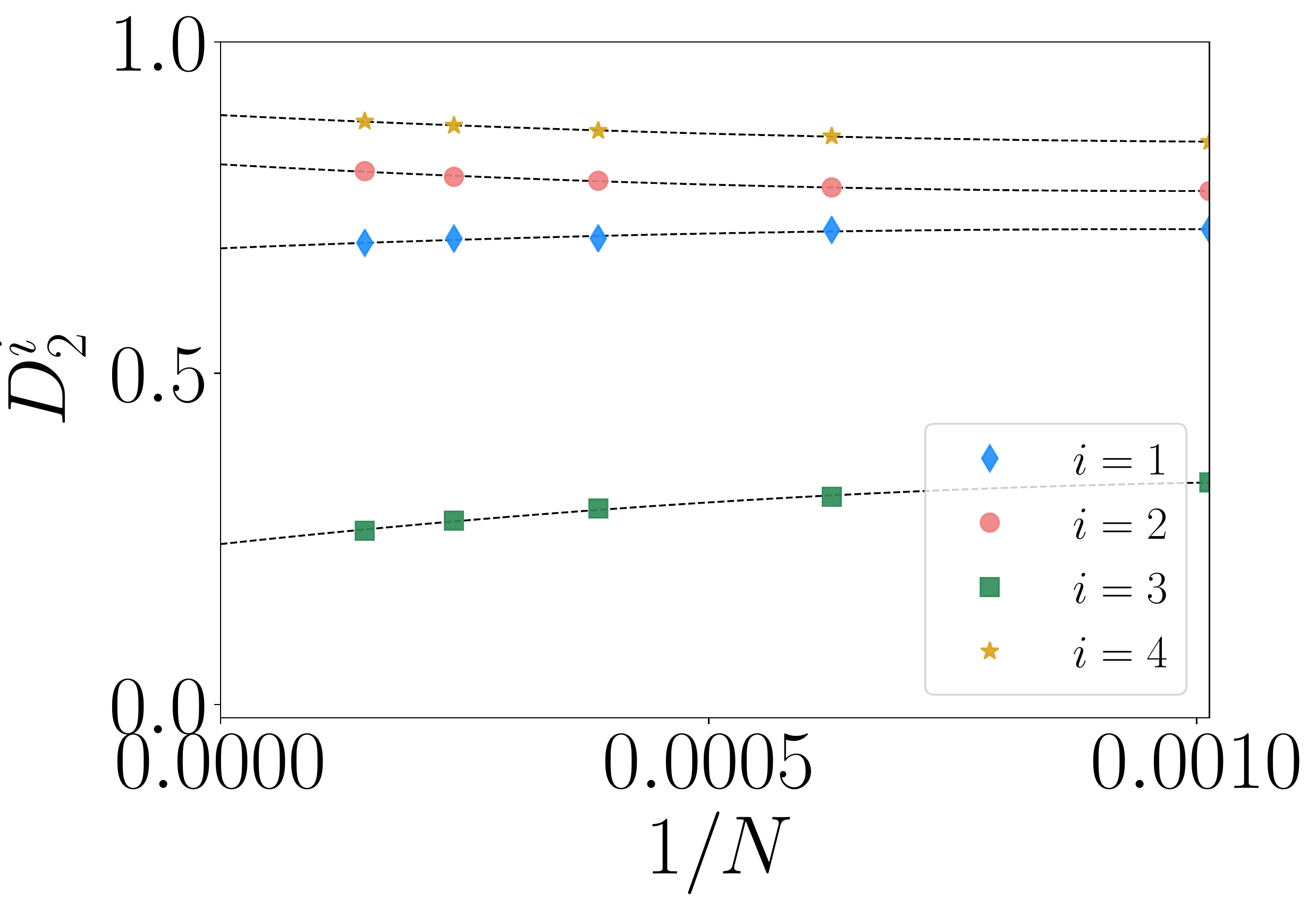}
\caption{\label{fig:scaling} \textbf{Scaling of the mean fractal dimension $\bar{D}_2$:} Scaling of the mean fractal dimension $\bar{D}_2$ for the eigenstates with positive energies $E/t$ of the Hamiltonian in eq.~\eqref{eq: hamiltonian} for $\Delta = 0.8$  $\alpha = 4.0$, $V/t = 2.0$ and $\xi \rightarrow \infty$ and for each of the regimes in $i = 1,2,3,4$ in Fig.~\ref{fig:histogram} (a).}
\end{figure}
In order to obtain the eigenstates of the Hamiltonian in eq.~\eqref{eq: hamiltonian} for $\xi \rightarrow \infty$ using exact diagonalization, we need to express in terms of the Bougolibov-de-Gennes (BdG) basis
\begin{equation}
\chi = \left(c_{0}, c^\dagger_{0}, c_{1}, c^\dagger_{1}, ..., c_{N-1}, c^\dagger_{N-1}\right)^T.
\end{equation}
Then,
\begin{equation}
    H = \chi^\dagger H_{BdG} \chi,
\end{equation}
where
\begin{equation}
H_{BdG} = 
\begin{pmatrix}
A_0 & B & C_2 &\cdots  & -B^\dagger\\
B^\dagger & A_1 & B & \cdots & C_{N-2}\\
\vdots  & \vdots  & \vdots  &  \ddots & \vdots\\
C_{N-2}^\dagger & C_{N-3}^\dagger & C_{N-4}^\dagger & \cdots & B\\
-B & C_{N-2}^\dagger & C_{N-3}^\dagger &  \cdots & A_{N-1},
\end{pmatrix}
\end{equation}
with $A_{i} = -V f(i) \sigma_z$, $B = \frac{t}{2}\sigma_z -\Delta i \sigma_y$ and $ C_l = -\frac{\Delta}{d_l^{\alpha}}i\sigma_y$. Here, we replace $l$ in eq.~\eqref{eq: hamiltonian} by $d_l = \text{min}(l, N-l)$ because we impose antiperiodic boundary conditions (APBC). In particular, the use of periodic boundary conditions can lead to cancelling terms such as $b_i b_j$ and $b_i b_{j+N}$ for Hamiltonians with long-range superconducting pairing. We can avoid this problem by instead using APBC~\cite{vodola14, lepori17}.
\section{\label{appendix:scaling} Scaling of the mean fractal dimension $\bar{D}_2$}

Figure~\ref{fig:scaling} shows the scaling of the mean fractal dimensions for the different regions indicated in Fig.~\ref{fig:histogram} (a). Region (1) corresponds to regular multifractal dimension with a mean fractal dimension converging to $\bar{D}_2^1 = 0.6$, while region (2) corresponds to a different multifractal behavior with a mean fractal dimension converging to a higher value, $\bar{D}_2^2 = 0.8$. Region (3) corresponds to localized states and region (4) corresponds to ergodic states. Note that the mean fractal dimension of the localized regime does not converge to zero in the thermodynamic limit, as expected. This is because for small values of the quasi-periodic potential, $V/t = 2.0$ in this case, we are very close to the other regimes and the localized states are still not fully localized in one site. Nevertheless, we note that the four regimes can be clearly distinguished in Fig.~\ref{fig:histogram} and Fig.~\ref{fig:scaling}.
%
%
%

%
%
\end{document}